\documentstyle[twoside]{article}

\catcode`\@=11
\long\def\@makefntext#1{
\protect\noindent \hbox to 3.2pt {\hskip-.9pt  
$^{{\eightrm\@thefnmark}}$\hfil}#1\hfill}		

\def\thefootnote{\fnsymbol{footnote}}
\def\@makefnmark{\hbox to 0pt{$^{\@thefnmark}$\hss}}	
	
\def\ps@myheadings{\let\@mkboth\@gobbletwo
\def\@oddhead{\hbox{}
\rightmark\hfil\eightrm\thepage}   
\def\@oddfoot{}\def\@evenhead{\eightrm\thepage\hfil
\leftmark\hbox{}}\def\@evenfoot{}
\def\sectionmark##1{}\def\subsectionmark##1{}}

\oddsidemargin=\evensidemargin
\renewcommand{\thefootnote}{\fnsymbol{footnote}}

\newcounter{sectionc}\newcounter{subsectionc}\newcounter{subsubsectionc}
\renewcommand{\section}[1] {\vspace{12pt}\addtocounter{sectionc}{1} 
\setcounter{subsectionc}{0}\setcounter{subsubsectionc}{0}\noindent 
	{\tenbf\thesectionc. #1}\par\vspace{5pt}}
\renewcommand{\subsection}[1] {\vspace{12pt}\addtocounter{subsectionc}{1} 
	\setcounter{subsubsectionc}{0}\noindent 
	{\bf\thesectionc.\thesubsectionc. {\kern1pt \bfit #1}}\par\vspace{5pt}}
\renewcommand{\subsubsection}[1] {\vspace{12pt}\addtocounter{subsubsectionc}{1}
	\noindent{\tenrm\thesectionc.\thesubsectionc.\thesubsubsectionc.
	{\kern1pt \tenit #1}}\par\vspace{5pt}}
\newcommand{\nonumsection}[1] {\vspace{12pt}\noindent{\tenbf #1}
	\par\vspace{5pt}}

\newcounter{appendixc}
\newcounter{subappendixc}[appendixc]
\newcounter{subsubappendixc}[subappendixc]
\renewcommand{\thesubappendixc}{\Alph{appendixc}.\arabic{subappendixc}}
\renewcommand{\thesubsubappendixc}
	{\Alph{appendixc}.\arabic{subappendixc}.\arabic{subsubappendixc}}

\renewcommand{\appendix}[1] {\vspace{12pt}
        \refstepcounter{appendixc}
        \setcounter{figure}{0}
        \setcounter{table}{0}
        \setcounter{lemma}{0}
        \setcounter{theorem}{0}
        \setcounter{corollary}{0}
        \setcounter{definition}{0}
        \setcounter{equation}{0}
        \renewcommand{\thefigure}{\Alph{appendixc}.\arabic{figure}}
        \renewcommand{\thetable}{\Alph{appendixc}.\arabic{table}}
        \renewcommand{\theappendixc}{\Alph{appendixc}}
        \renewcommand{\thelemma}{\Alph{appendixc}.\arabic{lemma}}
        \renewcommand{\thetheorem}{\Alph{appendixc}.\arabic{theorem}}
        \renewcommand{\thedefinition}{\Alph{appendixc}.\arabic{definition}}
        \renewcommand{\thecorollary}{\Alph{appendixc}.\arabic{corollary}}
        \renewcommand{\theequation}{\Alph{appendixc}.\arabic{equation}}
        \noindent{\tenbf Appendix \theappendixc #1}\par\vspace{5pt}}
\newcommand{\subappendix}[1] {\vspace{12pt}
        \refstepcounter{subappendixc}
        \noindent{\bf Appendix \thesubappendixc. {\kern1pt \bfit #1}}
	\par\vspace{5pt}}
\newcommand{\subsubappendix}[1] {\vspace{12pt}
        \refstepcounter{subsubappendixc}
        \noindent{\rm Appendix \thesubsubappendixc. {\kern1pt \tenit #1}}
	\par\vspace{5pt}}

\newcommand{\textlineskip}{\baselineskip=13pt}
\newcommand{\smalllineskip}{\baselineskip=10pt}

\def\eightcirc{
\begin{picture}(0,0)
\put(4.4,1.8){\circle{6.5}}
\end{picture}}
\def\eightcopyright{\eightcirc\kern2.7pt\hbox{\eightrm c}} 

\newcommand{\copyrightheading}[1]
	{\vspace*{-2.5cm}\smalllineskip{\flushleft
	{\footnotesize International Journal of Modern Physics A, #1}\\
	{\footnotesize $\eightcopyright$\, World Scientific Publishing
	 Company}\\
	 }}

\newcommand{\publisher}[2]{{\begin{center}\footnotesize\smalllineskip 
	Received #1\\
	Revised #2
	\end{center}
	}}

\def\abstracts#1#2#3{{
	\centering{\begin{minipage}{4.5in}\baselineskip=10pt\footnotesize
	\parindent=0pt #1\par 
	\parindent=15pt #2\par
	\parindent=15pt #3
	\end{minipage}}\par}} 

\renewenvironment{thebibliography}[1]
	{\frenchspacing
	 \ninerm\baselineskip=11pt
	 \begin{list}{\arabic{enumi}.}
	{\usecounter{enumi}\setlength{\parsep}{0pt}
	 \setlength{\leftmargin 12.7pt}{\rightmargin 0pt} 
	 \setlength{\itemsep}{0pt} \settowidth
	{\labelwidth}{#1.}\sloppy}}{\end{list}}

\newcounter{itemlistc}
\newcounter{romanlistc}
\newcounter{alphlistc}
\newcounter{arabiclistc}

\newcommand{\fcaption}[1]{
        \refstepcounter{figure}
        \setbox\@tempboxa = \hbox{\footnotesize Fig.~\thefigure. #1}
        \ifdim \wd\@tempboxa > 5in
           {\begin{center}
        \parbox{5in}{\footnotesize\smalllineskip Fig.~\thefigure. #1}
            \end{center}}
        \else
             {\begin{center}
             {\footnotesize Fig.~\thefigure. #1}
              \end{center}}
        \fi}

\newcommand{\tcaption}[1]{
        \refstepcounter{table}
        \setbox\@tempboxa = \hbox{\footnotesize Table~\thetable. #1}
        \ifdim \wd\@tempboxa > 5in
           {\begin{center}
        \parbox{5in}{\footnotesize\smalllineskip Table~\thetable. #1}
            \end{center}}
        \else
             {\begin{center}
             {\footnotesize Table~\thetable. #1}
              \end{center}}
        \fi}

\def\@citex[#1]#2{\if@filesw\immediate\write\@auxout
	{\string\citation{#2}}\fi
\def\@citea{}\@cite{\@for\@citeb:=#2\do
	{\@citea\def\@citea{,}\@ifundefined
	{b@\@citeb}{{\bf ?}\@warning
	{Citation `\@citeb' on page \thepage \space undefined}}
	{\csname b@\@citeb\endcsname}}}{#1}}

\newif\if@cghi
\def\cite{\@cghitrue\@ifnextchar [{\@tempswatrue
	\@citex}{\@tempswafalse\@citex[]}}
\def\citelow{\@cghifalse\@ifnextchar [{\@tempswatrue
	\@citex}{\@tempswafalse\@citex[]}}
\def\@cite#1#2{{$\null^{#1}$\if@tempswa\typeout
	{IJCGA warning: optional citation argument 
	ignored: `#2'} \fi}}

\def\pmb#1{\setbox0=\hbox{#1}
	\kern-.025em\copy0\kern-\wd0
	\kern.05em\copy0\kern-\wd0
	\kern-.025em\raise.0433em\box0}

\def\fnt#1#2{\footnotetext{\kern-.3em
	{$^{\mbox{\scriptsize #1}}$}{#2}}}

\def\fpage#1{\begingroup
\voffset=.3in
\thispagestyle{empty}\begin{table}[b]\centerline{\footnotesize #1}
	\end{table}\endgroup}

\def\runninghead#1#2{\pagestyle{myheadings}
\markboth{{\protect\footnotesize\it{\quad #1}}\hfill}
{\hfill{\protect\footnotesize\it{#2\quad}}}}
\font\tenrm=cmr10
\font\tenit=cmti10 
\font\tenbf=cmbx10
\font\bfit=cmbxti10 at 10pt
\font\ninerm=cmr9

\font\eightrm=cmr8

\def\qed{\hbox{${\vcenter{\vbox{			
   \hrule height 0.4pt\hbox{\vrule width 0.4pt height 6pt
   \kern5pt\vrule width 0.4pt}\hrule height 0.4pt}}}$}}

\renewcommand{\thefootnote}{\fnsymbol{footnote}}	

\begin{document}

\runninghead{Some Comments on Branes, G-flux, and K-theory
$\ldots$} {Some Comments on Branes, G-flux, and K-theory
 $\ldots$}

\normalsize\textlineskip
\thispagestyle{empty}
\setcounter{page}{1}

\copyrightheading{}			

\vspace*{0.88truein}

\fpage{1}
\centerline{\bf Some Comments on Branes, G-flux, }
\vspace*{0.035truein}
\centerline{\bf and K-theory }
\vspace*{0.37truein}
\centerline{\footnotesize Gregory Moore }
\vspace*{0.015truein}
\centerline{\footnotesize\it Department of Physics, 
Rutgers University,}
\baselineskip=10pt
\centerline{\footnotesize\it Piscataway, New Jersey, 08855}
\vspace*{10pt}
\vspace*{0.225truein}
\publisher{(received date)}{(revised date)}

\vspace*{0.21truein}
\abstracts{Summary of a talk explaining three ways in which 
string theory and M-theory are related to the mathematics of 
K-theory.}{}{}


\vspace*{1pt}\textlineskip	
\section{Introduction}	
\vspace*{-0.5pt}
\noindent
This talk was divided into 3 parts. In each part we describe how 
considerations related to   D-branes and RR fields of type II 
string theory naturally 
lead to mathematics related to K-theory. Part one reviews 
some work done with D.-E. Diaconescu and E. Witten, 
and reported in \cite{dmwi,dmwii}.  Here we review the 
way K-theory theta functions are related to M-theory partition 
functions. Part two reports on 
some work in progress with G. Segal. Details will appear in 
a future publication \cite{mooresegal}. 
Here we show how simple considerations in topological field 
theory lead to a picture of D-branes being classified by 
K-theory. 
Part three announced  material which was   done with J. Harvey, 
and has been described in detail  in 
\cite{hmk}. Here we explain how the K-theory of $C^*$ algebras 
fits in naturally with recent developments on tachyon condensation 
in the presence of B-fields. 

\pagebreak

\textheight=7.8truein
\setcounter{footnote}{0}
\renewcommand{\thefootnote}{\alph{footnote}}

\section{Partition functions: From M to K}
\noindent
The partition function $Z_{IIA}$ of type IIA superstring theory 
on a smooth compact spin  10-manifold $X$ should be related to the partition 
function $Z_M$ of $M$-theory on the associated manifold $Y= X \times S^1$. 
Nevertheless, showing that this is the case when $X$ is topologically 
complicated turns out to be a very subtle problem. This is 
the problem which was addressed and solved in detail in 
\cite{dmwi,dmwii}. 

In brief, we consider a long-distance/weak-coupling limit. In 
the IIA string we take $g_{\mu\nu} = t g_{\mu\nu}^0$, where 
$g_{\mu\nu}^0$ is some fixed metric and we let 
$t\to +\infty, g_{\rm string} \to 0$. On the $M$-theory 
side we take the long-distance limit (except for the $M$-theory 
circle!). In both theories the partition functions reduce to 
the form: 
\begin{equation}
Z=({\rm simple\quad factors})\times \sum_{{\rm G-flux}} e^{-S(G)}\label{genstruc}
\end{equation}
Formulating precisely both the sum and the measure in 
Eq.\ref{genstruc} is tricky. 

\subsection{Sum in type IIA theory}
\noindent
In this case $G=G_0 + G_2 + \cdots + G_{10}$ is the sum of 
all fluxes. In formulating the sum there are two basic 
imputs one must take into account. 
First, even in the IIA theory, one should consider $G$ to be a
{\it selfdual field}:  $G=*G$. Second,  there is a quantization 
condition on the allowed fluxes. In \cite{mw,freed} it is argued that 
the quantization condition is 
\begin{equation}
[{G\over 2\pi} ]= {\rm ch}(x + \theta/2) (\hat A(TX))^{1/2}\label{kquant}
\end{equation}
where $[G]$ is the DeRham cohomology class of the $G$-field 
(a real differential form) while $x,\theta$ are $K$-theory classes. 
$\theta$ is a   quantum shift, and {\it the topological 
sectors for RR fluxes are labelled by a K-theory class $x\in K^0(X)$. }
(For a recent discussion see \cite{freedii}.) 
The definition of the partition function 
of a self-dual scalar field in 2 dimensions can be 
generalized to higher dimensions. This was done in 
\cite{fivebrane} for self-dual 3-forms in the 5-brane, 
and generalized in \cite{selfduality} to the case of
10-dimensional RR fields. The procedure amounts to the  
 quantization of 
a certain principally polarized abelian variety. 
The underlying torus is 
$K^0(X)\otimes R/(K^0(X)/K^0(X)_{tors}) $  
which turns out to have a natural metric, complex 
structure, and positive integral $(1,1)$ form. 
In this way 
one arrives at the type IIA K-theory theta function 
\begin{equation}
\Theta_{IIA} =  \sum  e^{-KE(G)}e^{i \Phi(G)} \label{iiathet}
\end{equation}
Here the sum runs over all cohomological values of 
$G_0,G_2,G_4$ consistent with the {\it existence } 
of a K-theory class $x$ such that Eq. \ref{kquant} 
can hold. The existence of such an $x$ puts constraints on 
the possible cohomology class for $G_4$ (as described in 
section 2.3 below). 
The kinetic energy is
\begin{equation}
KE(G) = t^5 \parallel G_0 \parallel^2 + t^3 \parallel G_2 \parallel^2 + 
t \parallel G_4\parallel^2
\label{kineenr}
\end{equation} 
and corresponds to the standard supergravity action. Here $\parallel \cdot \parallel^2$
is the standard Hodge norm in the metric $g_{\mu\nu}^0$. On the other hand, 
the phase is rather subtle, and follows from 
the considerations of \cite{selfduality,dmwi,dmwii}. 
It has the structure: 
\begin{equation}
e^{i \Phi(G)}= \Omega(x)\exp[ 2\pi i \int_X (-{1\over 15} G_2^5 + {1\over 6} G_2^3 G_4+ \cdots)]
 \label{kphase}
\end{equation}
The terms in the exponential are new topological phases which must be 
included in the 10-dimensional supergravity action in order to match 
properly to M-theory on a circle. The function $\Omega(x)$ takes 
values $\pm 1$, and is based on a mod-two index. It can be thought 
of as $(-1)^{N(x)}$ where $N(x)$ is the number of Ramond fermion 
zeromodes on a IIB brane with K-theory charge $x$. 
One of the most striking aspects of $\Omega(x)$ is that there is 
no local formula for the mod two index\cite{asv} ! 
This is related to the well-known 
difficulties in formulating the action for a chiral field, 
and constitutes a significant departure from standard Lagrangian 
formulations of field theory. 
 
\subsection{M-Theory Partition Function}
\noindent
Now we define precisely the M-theory partition function on 
a compact smooth spin 11-manifold $Y$ in the long distance 
limit. 
As with the RR partition function, there is a subtle quantization 
condition and a phase. Both were analyzed by Witten a few 
years ago in 
\cite{fourflux}. 
The quantization condition depends on a {\it cohomology} class 
$a\in H^4(Y,Z)$, and is $[{G(a)\over 2\pi}] = a- \lambda/2$ where 
$\lambda$ is the characteristic class for the spin structure on 
$Y$. It satisfies $2\lambda = p_1(TY)$. The phase is, 
roughly speaking $\Omega_M(C) = \exp[2\pi i \int_Y (CGG + C X_8)]$, 
where $X_8$ is the famous $R^4$ correction to 11-dimensional 
supergravity \cite{duff,vw}. However, if $a\not=0$ then $C$ is not globally 
defined, and one must formulate this phase carefully. 
One approach (adopted in \cite{fourflux}) is to choose a bounding 
12-manifold $Z$,  $\partial Z = Y$ on which $G$ extends,  and set 
$\Omega_M(C) = \exp[2\pi i \int_Z (\alpha G^3 +\beta  G X_8)]$, 
where $\alpha, \beta $ are certain numerical coefficients. 
(That such extensions even exist is a nontrivial result of 
Stong.) 
While this is perfectly correct, it can be difficult to work with. 
Moreover, it is not manifestly well-defined because there can 
be different choices for $Z$. The best way to understand 
why the phase is well-defined, as pointed out by Witten in 
\cite{fourflux}, is to give the formulation of the phase in 
terms of $E_8$ gauge theory in 12 dimensions. 

Here we will give a slightly different
(but equivalent)  definition of the M-theory phase
based on work in progress with Emanuel Diaconescu
\cite{diacmoore}. 
An M-theory 3-form ``$C$-field'' can be defined to be a quadruple: 
$(V, A, G, c)$ where $V\to X$ is an $E_8$ vector bundle and $A$ is a 
connection on $V$. $G\in \Omega^{4}(X,R)$ is a real differential 
form,  and $c\in \Omega^3(X,R)/\Omega^3_Z(X)$, where 
$\Omega^3_Z(X)$ are 3-forms with integral periods (they are 
necessarily closed). These data must satisfy 
\begin{equation}
{G\over 2\pi} =   {1\over 60} {\rm Tr}_{248} {F^2\over 8\pi^2}  +{1\over 32 \pi^2}
{\rm Tr}{R^2} + d{c \over 2 \pi} 
 \label{constraint}
\end{equation}
and are subject to   an equivalence relation 
$(V_1, A_1, G_1, c_1) \sim
(V_2, A_2,  G_2 , c_2)$ if $G_2=G_1$, and
 $CS(A_1,A_2) = c_2-c_1$, where $CS$ is the Chern-Simons 
form associated to the above characteristic class. 
It should be stressed that this definition is equivalent 
to the statement that the M-theory 3-form is a Cheeger-Simons 
differential character $C \in \widehat{H}^3(Y,U(1))$. The 
proof of this fact relies on the remarkable result that 
$E_8$ bundles in less than 15 dimensions are classified by 
a single characteristic class $a\in H^4(Y,Z)$ \cite{wittensb}. 
Let us denote the corresponding bundle by $V(a)$. 

In terms of these data the M-theory phase is given by 
\begin{equation}
\Omega_M(C)
= \exp\Biggl[
2\pi i \biggl( {\eta(D_{V(a)}) + h(D_{V(a)}) \over 4}
 +   {\eta(D_{RS}) + h(D_{RS}) \over 8}
 \biggr) \Biggr]  \cdot \omega(c) 
 \label{mphase}
\end{equation}
Here $D_V$ is the Dirac operator coupled to the connection $A$ on the 
bundle $V(a)$,  $D_{RS}$ is the Rarita-Schwinger operator, 
 $h(D)$ = number of zeromodes of the operator $D$ on Y,  and $\eta(D)$ is the 
 eta invariant of Atiyah-Patodi-Singer.  The phase $\omega(c)$ 
is given by 
\begin{equation}
\omega(c)=
\exp\Biggl[
2\pi i \int_Y\biggl( c( \bar G^2 +X_8) + cdc \bar G +{1\over 3}c (dc)^2 
\biggr) \Biggr]  
\label{littlec}
\end{equation}
where $\bar G = {1\over 60} {\rm Tr}_{248} {F^2\over 8\pi^2}  +{1\over 32 \pi^2}
{\rm Tr}{R^2}$. The generalization of this formula to 11-manifolds 
with boundary is not trivial, and will be reported in \cite{diacmoore}. 
\footnote{The formula for $\omega(c)$ reported in the talk at   Strings2000 
omitted the second and third terms above.} 

\subsection{Equality of the partition functions} 

In the case when $Y=X\times S^1$, with the supersymmetric spin structure on the 
circle $S^1$, and a ``$C$-field'' pulled back from $X$ we can compare the 
partition functions at leading order, ${\cal O}(e^{-t})$ in the large 
distance expansion. Such contributions come solely from $G_4$. 
In \cite{dmwi,dmwii} it is shown that both partition 
functions reduce to determinants times the theta function 
\begin{equation}
\Theta = \sum e^{-\parallel G(a) \parallel^2} (-1)^{f(a)}
\label{finalthet} 
\end{equation} 
where the sum is over cohomology classes $a\in H^4(X,Z)$ such that 
$Sq^3(a)=0$, where $Sq^3$ is the Steenrod squaring operation. 
On the IIA side, this is precisely the condition for the existence 
of a K-theory class $x$ with ${\rm ch}(x)=a+ \cdots$. On the 
M-theory side, this arises because the subtle M-theory 
phase is sensitive to torsion information in the cohomology class 
$a$ of the M-theory 4-form. The sum over the torsion classes 
projects onto $a$'s such that $Sq^3(a)=0$. It is in this somewhat 
indirect way that the classification of IIA  RR fluxes via K-theory 
turns out to be compatible with the classification of M-theory 
4-flux by cohomology. The $Z/2Z$-valued 
function $f(a)$ is a mod-two index for the Dirac operator 
coupled to an $E_8$ connection in 10-dimensions, and again 
follows, nontrivially, from the definition of the phase in 
both M-theory and in IIA theory. 

\subsection{Further results} 

The basic computation sketched above can be extended in many 
directions. First, as shown in \cite{dmwi,dmwii} the result 
extends to the next subleading order in the large $t$ expansion 
at order $e^{-t^3}$. This involves summing over nontrivial 
circle bundles $Y \to X$ on the M-theory side, and correspondingly 
over nontrivial configurations for $G_2$ as well as for $G_4$ 
on the IIA side. There are many directions for further work 
including computation of one-loop determinants, extension to the 
type I setting, and inclusion of topologically nontrivial B-fields. 
\footnote{The considerations of section 2.1 extend straightforwardly 
to the case of flat $B$-fields $B\in H^2(X,U(1))$, if $[H]\in H^3(X,Z)$ 
is torsion. 
They reproduce rather nicely the couplings of the RR sector 
to the $B$-field.} Work in progress with E. Diaconescu 
\cite{diacmoore} is extending 
the result to include ``instanton'' amplitudes where membranes and 
fivebranes are inserted in the partition function. 

Among the many open problems in the subject two are outstanding, 
but probably very difficult. First,  
$E_8$ gauge theory has thus far proved to be a mathematical 
convenience. Whether or not there is a deeper physical 
significance to the presence of 11 and 12-dimensional 
$E_8$ gauge theories is a fascinating open issue. 
Second, it is clear that the relation of K-theory to 
RR charges should have a profound generalization to include 
fundamental string and NS 5-brane charges. One focal point 
for finding this generalization is the issue of how 
the relation of RR charges and K-theory can be compatible with 
the S-duality of IIB string theory. (This was, in fact, 
the central motivating question behind \cite{dmwi,dmwii}.) 
For some provocative recent results on this subject 
see \cite{Aspinwall}.

\section{Sewing constraints and D-branes}

The previous section focused on a long-distance approach to 
the emergence of K-theory. It is of interest to understand how 
the relevance of K-theory can be seen from a more ``microscopic'' 
or fundamental approach, for example, from considerations of 
conformal field theory, or of string field theory. Such 
considerations lead one directly to think about the K-theory 
of algebras, a subject that was very much ``in the air'' at 
the Strings2000 meeting. In this section we report on one way 
of understanding the connection between K-theory and D-branes 
based on simple considerations of sewing and topological field 
theory. This is based on on-going work with Graeme Segal. 
We recommend Segal's Stanford lectures \cite{segalstanford} 
as useful background.  

The axiomatic formulation of open and closed 
strings was   considered some time ago 
by P. Horava \cite{horava}. Nevertheless, we think it 
is interesting to reformulate it in the light of the 
connection between K-theory and D-branes. 
After submission of this manuscript to the editors, 
but before putting it on hep-th, a paper appeared with 
some overlapping results for this section \cite{lazaroiu}.

Given a closed string background one can ask: ``What are the 
possible D-branes?'' Similarly, given a closed conformal 
field theory ${\cal C}$ one can ask: ``What are the possible 
associated boundary conformal field theories?'' These questions 
are too hard to answer at present. 
Nevertheless, it turns out that if one replaces a conformal 
field theory by a 2d topological field theory then the question 
is solvable, yet not entirely trivial. 

Recall the ancient folktheorem that 2D topological field theories 
are in 1-1 correspondence with commutative Frobenius algebras: 
In/out circles map to in/out Hilbert spaces, and an oriented surface 
maps to a linear operator between in and out spaces. The multiplication 
in the Frobenius algebra
is defined by the pants diagram, the trace and unit are 
defined by the operator corresponding to the disk, with 
opposite orientations.  The axioms of a commutative associative 
algebra are equivalent to the consistency of sewing arbitrary 
surfaces from these three components. Let us now allow both 
open and closed strings. Then, the 2 dimensional surfaces have 
two kinds of boundaries. There are in- and out-going intervals 
corresponding to the in- and out-going open strings, and there 
are the ``free boundaries'' corresponding to the ends of the 
open string ``moving along a D-brane.'' Free boundaries 
carry boundary condition labels $a,b,\dots$ which should be thought of as 
objects in a linear category \cite{segalstanford}. 
The open string sectors ${\cal O}_{ab}$ are the morphisms between 
the objects $a$ and $b$ in the linear category. Our 
question: ``Given a closed 2d topological field theory, what 
are the possible boundary conditions?''  should be asked as 
two questions: First, ``What are the algebraic conditions that 
encode consistency of open and closed string sewing?'' and 
second, ``What linear categories are consistent with these 
conditions?''  It is useful to focus first on the case of a 
single boundary condition on both ends of the open string, 
that is, to let ${\cal O} = {\cal O}_{aa}$. Then we have 

\vspace*{12pt}
\noindent
{\bf Theorem 1:} To give an open and closed topological field theory 
is to give 

\def\CC{{\cal C}} 
\def\CO{{\cal O}}

1. A commutative Frobenius algebra $(\CC, \theta_{\CC},1_{\CC})$. 

2. A not necessarily commutative Frobenius algebra $(\CO, \theta_{\CO}, 1_{\CO})$. 

3. A homomorphism $\iota_*:\CC \to Z(\CO)$, where $Z(\CO)$ is
 the center of $\CO$, such that 
$\iota_*(1_{\CO})= 1_{\CC}$, and such that: 
\begin{equation}
\pi = \iota_* \iota^*
\label{cardy}
\end{equation}
Here $\iota^*$ is the adjoint to $\iota_*$, defined by 
$\theta_{\CO}(\psi \iota_*(\phi)) = \theta_{\CC}(\iota^*(\psi)\phi)$, 
while $\pi:\CO \to \CO$ is an operator defined by the double-twist 
diagram. 
\vspace*{12pt}

The condition Eq.\ref{cardy} is closely related to the 
``Cardy condition'' of boundary conformal field theory. 
The operator $\pi$ is defined 
using only the open string data $(\CO, \theta_{\CO}, 1_{\CO})$. 
Indeed, if $\psi_\mu$ is a basis for $\CO$ and 
$\psi^\mu$ is a dual basis relative to $\theta_{\CO}$ then 
$\pi(\psi) = \sum_\mu \psi_\mu \psi \psi^\mu$. In pictures, 
Eq. \ref{cardy} follows because the double twist diagram 
can be viewed both as an open string diagram, and also as 
a closed string channel diagram. Indeed $\iota^*$ is the open 
to closed string transition, while $\iota_*$ is the 
closed to open transition. 

The claim of the theorem is that in the case of open and 
closed strings with a single type of boundary condition 
on all free boundaries, the above conditions are equivalent 
to the consistency conditions for sewing. This statement 
is not trivial to prove. The conditions are roughly 
the same as those found in \cite{lewellen} but are 
not precisely the same. 

Let us now consider the solutions to the above 
conditions. In general this is a difficult problem,
but if the ``fusion rules'' (i.e. the regular representation 
matrices) of $\CC$ are diagonalizable then we can 
classify the $\CO$'s. When the fusion rules are 
diagonalizable we say $\CC$ is ``semisimple.'' In 
this case we may introduce basic idempotents 
such that $\epsilon_i \epsilon_j = \delta_{ij} \epsilon_i$ 
and $\CC = \oplus_i C\cdot \epsilon_i$. Indeed, given 
any basis $\phi_\mu$ for $\CC$ one can diagonalize the 
fusion rules with a matrix $S_\mu^i$. Letting $\mu=0$ 
correspond to the identity element we can write: 
$\epsilon_i = \sum_\mu S_0^i (S^{-1})_i^\mu \phi_\mu$. 

\vspace*{12pt}
\noindent
{\bf Theorem 2:} If $\CC$ is semisimple then 
$\CO  = {\rm End}_{\CC}(M)$ for $M$ a finitely 
generated projective $\CC$-module. 
\vspace*{12pt}
\noindent

The theorem states that the possible $\CO$'s are simply 
sections of a vector bundle over the 
``spacetime'' ${\rm Spec}(\CC)$. 
One should think of $\epsilon_i$ as corresponding to the 
spacetime points in ${\rm Spec}(\CC)$. Recall that in 
the Gelfand-Naimark theorem one defines ${\rm Spec}(\CC)$ 
as the space of characters of the algebra $\CC$. Indeed, the 
character corresponding to $\epsilon_i$ 
is $\chi(\phi) = \theta_{\CC}(\epsilon_i \phi)/\theta_i$
where 
$\theta_i = \theta_{\CC}(\epsilon_i)$. These traces 
are part of the invariants of a Frobenius algebra.
Then $\CO = \oplus_i {\rm End}(W_i)$ for some collection 
of finite dimensional vector spaces $W_i$. 

In this formalism one can easily work out the formula 
for the ``boundary state,'' defined by $B=\iota^*(1_{\CO})$. 
This object is an element of $\CC$ and has the 
property that insertions of $n$ factors of $B$ 
in $\theta_{\CC}(B^n \cdots)$ corresponds geometrically 
to adding $n$ holes to the surface. In formulae 
$B= \sum_i {\rm dim}(W_i)\epsilon_i/\sqrt{\theta_i}$, 
  The 
squareroot in this formula is significant and is 
related to the standard fact that the closed string 
coupling is the square of the open string coupling. 
Moreover,   if we consider a 
{\it family} of Frobenius algebras, then, if the subfamily  
of semisimple algebras has nontrivial fundamental 
group, transport around nontrivial 
loops can result in  monodromy such that $\epsilon_i \to \epsilon_{\sigma(i)}$ 
where $\sigma$ is a permutation, while $\sqrt{\theta_i} 
\to \pm \sqrt{\theta_{\sigma(i)}}$. Because of this one 
should generalize vector spaces to virtual vector spaces above 
to allow for negative integers in $\dim(W_i)$. 

Now, let us consider multiple boundary conditions. The 
open string spaces $\CO_{ab}$ for distinct boundary conditions 
$a,b$ is a bimodule for $\CO_{aa} \otimes \CO_{bb}$. 
The Cardy condition generalizes in the obvious 
way $\pi_{a}^b = \iota_a \iota^b$, where $\iota^b$ is the 
open to closed transition for boundary conditions of 
type bb, and $\iota_a$ is the closed to open transition 
for boundary conditions of type aa. This condition 
can be shown to imply that $\CO_{ab} = {\rm Hom}(W_a, W_b)$
where $W_a, W_b$ are projective modules for $\CC$. 
The conclusion then is that {\it the linear category classifying 
the boundary conditions for the closed 2d topological field theory 
is the $K$-theory $K_0(\CC)$ for the commutative closed string 
Frobenius algebra.} 

Many interesting examples of the above theorems applied to 
families of Frobenius algebras can be given. Moreover, the 
discussion can be   generalized to ``orbifolds'' that is, 
to the equivariant case. Analogs of theorems 1 and 2 above
can be stated, at least in the semisimple case. 
It is possible to introduce a ``$B$-field'' 
even in this topological setting. The ``untwisted sector'' 
of the theory defines an ordinary commutative Frobenius 
algebra $\CC_1$, and ${\rm Spec}(\CC_1)$ will be a finite 
$G$-space $X$, where $G$ is the orbifold group. 
The $B$-field is valued in $G$-equivariant 
cohomology $H^2_G(X;U(1))$ 
and has a corresponding fieldstrength $h\in H^3_G(X;Z)$. 
 The linear category is then 
given by the $K$-theory $K_{G,h}(X)$ 
of $G$-equivariant, twisted bundles 
of algebras over the ``spacetime'' $X$, in accord with 
previous works. Details   will be described in \cite{mooresegal}. 

The above results might have some bearing on the subject
of topological open strings and D-branes in Calabi-Yau 
manifolds. For recent work on this subject see 
\cite{Hori,Douglas,Lazaroiuii}.

\section{Noncommutative Tachyons and K-theory} 

Part three of the talk announced results that have since 
been published in \cite{hmk}. We will therefore be 
very brief here. 

There has been much recent progress on understanding tachyon 
condensation by using the technology of noncommutative field 
theory   \cite{gms,hklm,dmr,wittenlennyfest,wittenstrings}. Consider the 
bosonic or the type II string. The general 
picture is to consider spacetime to be a product of a commutative 
and a noncommutative manifold $X_c \times X_{nc}$. A field on 
a noncommutative manifold is equivalent to an operator on 
Hilbert space. Therefore,  if an unstable D25, or D9 
brane wraps this spacetime then its tachyon field will 
be described by a field on $X_c$, valued in operators on 
Hilbert space. At large noncommutativity parameter the 
equations of motion for the tachyon field say that it is 
a projection operator (in the bosonic string) or a partial 
isometry (in the type II string). This fits in perfectly 
with the K-theory classification of D-branes. Indeed, 
the isomorphism classes of complex rank $n$ vector 
bundles on $X_c$ are in one-one correspondence with 
homotopy classes $[X_c, BU(n)]$ where $BU(n)$ is the 
space of rank $n$ projection operators on Hilbert space. 
Similarly, the Atiyah-Janich model for K-theory shows 
that we can identify $K^0(X_c)$ with the homotopy 
classes $[X,{\cal F}]$ where ${\cal F}$ is the space 
of Fredholm operators on Hilbert space. By polar 
decomposition one 
can restrict to the space of partial isometries. 

In the type II string a very special class of partial 
isometries can be written corresponding to the 
``noncommutative Atiyah-Bott-Shapiro construction.'' 
One takes $[x^i, x^j] = - i \theta^{ij}$ with $\theta$ 
of maximal rank and $i=1,\dots, 2p$. Letting $\Gamma_i$ 
be chiral gamma matrices of rank $2^{p-1}$ we can form 
the partial isometry in the polar decomposition of 
$\Gamma_i x^i$. This is the tachyon field corresponding 
to 
condensing a D9 brane to a D(9-2p) brane transverse to 
a noncommutative plane of dimension 2p. The fact that 
the D-brane charge is unchanged by turning on a B-field 
is equivalent to the index theorem identifying topological 
and analytical indices. For further details, and 
further developments of these ideas, see \cite{hmk}, 
and references therein. 

Recently, in collaboration with E. Martinec 
we have generalized these ideas to include 
D-branes on orbifolds. The algebra of functions 
on $X_{nc}$ is replaced by a crossed-product 
algebra. Fractional branes and discrete torsion 
are very naturally and easily incorporated 
into the formalism. In addition, the formalism 
gives an interesting perspective on the  formulation of the 
theory of D-branes in asymmetric orbifolds. 
Details will appear in 
\cite{martinecmoore}. 

\nonumsection{Acknowledgments}
\noindent
I would like to thank the organizers for the 
opportunity to present this talk at Strings2000. 
The above work is all collaborative and would
never have been possible without the insights 
and contributions of D.-E. Diaconescu, J. Harvey, 
E. Martinec, G. Segal, 
and E. Witten.  This work is supported by 
DOE grant DE-FG02-96ER40949

\nonumsection{References}

\end{document}